\documentclass[aps,pre,reprint,amssymb,amsmath,superscriptaddress,groupedaddress]{revtex4-1}
\usepackage{graphicx}
\usepackage{dcolumn}
\usepackage{bm}
\usepackage{subfigure}

\begin{document}

\title{Two Scenarios for the Onset and Suppression of Collective Oscillations in Heterogeneous Populations of Active Rotators}

\author{Vladimir Klinshov}
\email{vladimir.klinshov@ipfran.ru}
\affiliation{Institute of Applied Physics of the Russian Academy of Sciences, 46 Ulyanov Street,
603950 Nizhny Novgorod, Russia}
\author{Igor Franovi\'c}
\email{franovic@ipb.ac.rs}
\affiliation{Scientific Computing Laboratory, Center for the Study of Complex Systems,
Institute of Physics Belgrade, University of Belgrade, Pregrevica 118, 11080 Belgrade, Serbia}

\date{\today}
\begin{abstract}
We consider the macroscopic regimes and the scenarios for the onset and the suppression of collective oscillations in a heterogeneous population of active rotators, comprised of excitable or oscillatory elements. We analyze the system in the continuum limit within the framework of Ott-Antonsen reduction method, determining the states with a constant mean field and their stability boundaries in terms of the characteristics of the rotators’ frequency distribution. The system is established to display three macroscopic regimes, namely the homogeneous stationary state, the oscillatory state and the heterogeneous stationary state, whereby the transitions between the characteristic domains involve a complex bifurcation structure, organized around three codimension-two bifurcation points: a Bogdanov-Takens point, a cusp point and a fold-homoclinic point. Apart from the monostable domains, our study also reveals two domains admitting bistable behavior, manifested as coexistence between the two stationary solutions, or between a stationary and a periodic solution. It is shown that the collective mode may emerge via two generic scenarios, guided by a SNIPER or the Hopf bifurcation, such that the transition from the homogeneous to the heterogeneous stationary state under increasing diversity may follow the classical paradigm, but may also be hysteretic. We demonstrate that the basic bifurcation structure holds qualitatively in presence of small noise or small coupling delay, with the boundaries of the characteristic domains shifted compared to the noiseless and delay-free case.
\end{abstract}


\maketitle

\section{Introduction} \label{intro}

The onset of a collective mode mediated via a transition to synchrony is a fundamental paradigm of macroscopic behavior in a broad variety of fields, ranging from neuroscience and other biologically-inspired models to chemistry, technology and social science \cite{PRK03,ABV05}. A classical approach within the theory of nonlinear dynamics is to regard populations exhibiting a collective mode as macroscopic oscillators \cite{BRP09,FTVB13,OLPT10}, which can then interact with other populations or be subjected to external stimuli. In this context, we investigate an important problem of the emergence and the suppression of collective oscillations in populations comprised of units with nonuniform intrinsic parameters, which are drawn from a certain probability distribution. Such nonuniformity is a manifestation of variability \cite{S00,KT10,TSTC07,PM06}, a ubiquitous feature that often makes it more realistic to consider heterogeneous rather than homogeneous assemblies. Depending on the particular application, variability may alternatively be referred to as diversity, heterogeneity, impurities, or quenched noise. In many cases, the diversity can be large enough to give rise to qualitative differences in individual dynamics of units, such that some of the active elements within a population may be self-oscillating while the others are excitable.

The classical Kuramoto paradigm \cite{K84} addresses the scenario where the diversity is manifested at the quantitative level alone, since all the units are considered to be self-oscillating. There, the continuous transition to synchrony occurs once the coupling between the oscillators becomes strong enough to overcome the effects of diversity \cite{ABV05,RPJK16}. Nevertheless, the diversity alone has been shown to be capable, under appropriate conditions, to enhance the response of an assembly to external forcing or to promote synchronization \cite{TSTC07, KT10,TMTG06}. Moreover, in the case of heterogeneous assemblies made up of excitable and oscillatory units rather than the oscillators alone, it has been demonstrated that the transition to synchrony with increasing diversity may be classical or reentrant, depending on the particular form of the units’ frequency distribution \cite{LCT10}. For such a setup, it has also been indicated that the collective firing emerges via a generic mechanism where the entrainment of units is degraded by increasing diversity \cite{TSTC07}.

In the present paper, we investigate the regimes of macroscopic behavior, as well as the scenarios for the onset and the suppression of collective oscillations in a heterogeneous population made up of oscillatory and excitable units, considering a model of active rotators with global sine coupling. Our analysis relies on the Ott-Antonsen reduction method \cite{OA08,OA09}, based on the Ansatz that the long-term macroscopic dynamics of such systems settles on a particular invariant attractive manifold. We first provide an exact description of macroscopic stationary states featuring a \emph{constant mean-field}, and then determine the bifurcations that outline the stability boundaries of the characteristic domains. While the stationary states and the associated self-consistency equation are obtained for an arbitrary distribution of natural frequencies, the subsequent bifurcation analysis is carried out for a \emph{uniform} frequency distribution on a bounded interval, which has the advantage of allowing for analytical tractability. We establish the complete bifurcation structure and demonstrate two generic scenarios for the emergence and the suppression of the collective mode. While the scenario featuring the successive onset and suppression of oscillations under increasing diversity has earlier been reported to be universal for heterogeneous populations with various distributions of the units' frequencies \cite{TMTG06,LCT10}, the other scenario, which involves a hysteretic behavior due to existence of bistability regions, is reported here for the first time.

Apart from diversity, the two additional ingredients influencing the dynamics in neuronal and other biophysical systems are coupling delays and noise \cite{TP01,PJ08,ZLS16}. In particular, realistic models often  have to include explicit coupling delays in order to describe the effects of finite velocity of signal propagation or the latency in information processing \cite{PJ08,A10,SHHD09,TFE00,RP04,PYP10}. On the other hand, creating coarse-grained models inevitably requires one to incorporate different sources of noise \cite{LGNS04,KF15,FK18,FK16,FK18a,FTPVB15,FPTKB15,BKNPF18}. Both coupling delay and noise may play an important role in the collective dynamics of a population. For example, in systems consisting just of excitable units, it is well known that the noise may play a constructive role, contributing to the onset of collective firing via synchronization of local noise-induced oscillations \cite{KWS98,A01,PHYS99,ZNFS03}. Concerning the effect of coupling delays, the standard Kuramoto model with uniform delays has been shown to exhibit the discontinuous rather than the continuous transition between the incoherent and coherent states, further having the synchronization frequency suppressed by the delay \cite{RPJK16,CKKH00}.

Our study evinces the robustness of the general physical picture, inherited from the noiseless and the delay-free case, in presence of \emph{small} coupling delay and \emph{small} noise. While the impact of small delay may be analyzed within the local stability approach we developed, the Ott-Antonsen method in principle does not allow one to treat stochastic assemblies. Only quite recently, an approach involving the so-called circular cumulants \cite{TGKP18,GP18} has been developed to incorporate a first-order correction to the Ott-Antonsen theory, which accommodates for the effects of noise. We perform numerical analysis of the system dynamics in presence of small noise and complement it with qualitative arguments.

The paper is organized as follows. In Section \ref{sec2}, we present the details of the model and provide the continuum limit formulation for the delay- and the noise-free setup, obtaining the Ott-Antonsen equation for the local order parameter. Section \ref{sec3} comprises the analytical results on the local structure of the macroscopic stationary states and the related self-consistency equation, derived for an arbitrary frequency distribution. In Section \ref{sec4}, the stability and bifurcation analysis of the stationary states is carried out for a particular distribution of frequencies, comparing the stability boundaries of the characteristic domains to those obtained in numerical experiments. In Section \ref{sec5}, it is shown that the basic bifurcation scenario persists in presence of small noise or small coupling delay. Section \ref{sec6} contains our concluding remarks.

\section{Model dynamics and the continuum limit formulation}
\label{sec2}

We consider a heterogeneous assembly of $N$ globally coupled active rotators described by:
\begin{align}
\dot{\theta_i}(t)&=\omega_i-a\sin{\theta_i}(t)-\frac{K}{N}\sum\limits_j\sin\left(\theta_i(t)-\right. \nonumber \\
&\left.\theta_j(t-\tau)+\alpha\right)+\sigma\eta_i(t), i=1,\dots N \label{eq1}
\end{align}
where the phase variables are $\theta_i\in S^1$, and the local dynamics is governed by the non-isochronicity parameter $a$ and the natural frequency $\omega_i$. Regarding the term "natural frequency", note that it will be used for convenience to describe the intrinsic parameter involving the quenched randomness, even though some units may exhibit excitable, rather than oscillatory behavior. The frequencies are distributed according to the probability density function $g(\omega)$ that satisfies $\int_{-\infty}^{\infty} g(\omega)d\omega=1$, and is characterized by the mean value $\Omega$ and the width $\Delta$, which we here explicitly refer to as the \emph{diversity} parameter. The individual unit rotates uniformly with the frequency $\omega_i$ for $a = 0$ only, whereas for $a>0$ its rotation becomes non-uniform, having the rotation direction dependent on the sign of $\omega_i$. The relation between $\omega_i$ and the parameter $a$ underlies the excitability feature of autonomous dynamics. In particular, $\omega_i$ constitutes the bifurcation parameter, such that for fixed $a$, an isolated unit lies in the excitable regime if $|\omega_i|<a$. In this case, the unit possesses a stable node, whereas the characteristic nonlinear threshold-like response is mediated by an unstable steady state. At $|\omega_i|=a$, an isolated unit undergoes a SNIPER bifurcation toward the oscillatory regime. The interactions are assumed to be uniform across the population, and are characterized by the coupling strength $K$, the coupling phase-lag $\alpha$ and the coupling delay $\tau$. The effect of random fluctuations is represented by the white Gaussian random forces $\eta_i$ of intensity $\sigma^2$, which act independently on each unit
($\langle \eta_i(t)\rangle=0, \langle \eta_i(t)\eta_j(t’)\rangle=\delta_{ij}\delta(t-t’)$).

As already indicated, in this and the following section we apply the Ott-Antonsen framework \cite{OA08,OA09} to investigate the collective dynamics of an heterogeneous assembly of active rotators in the delay- and the noise-free case $\tau=\sigma=0$. To this end, let us introduce the Kuramoto complex order parameter, which represents the center of mass of all rotators:
\begin{equation}
R(t)=\rho(t)e^{i\psi(t)}=\frac{1}{N}\sum\limits_je^{i\theta_j(t)}, \label{eq2}
\end{equation}
such that \eqref{eq1} can be rewritten as
\begin{equation}
\dot{\theta_i}=\omega_i-\frac{a}{2i}(e^{i\theta_i}-e^{-i\theta_i})+\frac{K}{2i}
(Re^{-i(\theta_i+\alpha)}-\overline{R}e^{i(\theta_i+\alpha)}), \label{eq3}
\end{equation}
where the bar denotes the complex conjugate. In the thermodynamic limit $N\rightarrow\infty$, the macroscopic state of the system can be described by the probability density function $f(\theta,\omega,t)$, which, for the considered moment $t$, gives the relative number of oscillators whose phases and frequencies are $\theta_i(t)\approx \theta$, $\omega_k\approx\omega$. The normalization condition required for the probability density function is $\int_0^{2\pi}f(\theta,\omega,t)d\theta=g(\omega)$. Given the conservation of oscillators, $f(\theta,\omega,t)$ has to fulfill the continuity equation
\begin{equation}
\frac{\partial f}{\partial t}+\frac{\partial}{\partial \theta}(fv)=0, \label{eq4}
\end{equation}
where the velocity is just
\begin{equation}
v(\theta,\omega,t)=\omega-\frac{a}{2i}(e^{i\theta}-e^{-i\theta})+\frac{K}{2i}
(Re^{-i(\theta+\alpha)}-\overline{R}e^{i(\theta+\alpha)}). \label{eq5}
\end{equation}
In the last expression, we have used the form of the Kuramoto mean field in the thermodynamic limit $N\rightarrow\infty$
\begin{equation}
R(t)=\int_{-\infty}^{\infty}d\omega\int_0^{2\pi}f(\theta,\omega,t)e^{i\theta}d\theta,\label{eq6}
\end{equation}
According to the Ott-Antonsen Ansatz \cite{OA08,OA09}, the long-term dynamics of the continuity equation \eqref{eq8} settles on a particular manifold of the form
\begin{equation}
f(\theta,\omega,t)=\frac{g(\omega)}{2\pi} \left( 1 + \sum_{n=1}^\infty
\left[ \overline{z}^n(\omega,t) e^{i n \theta} + z^n(\omega,t) e^{-i n \theta} \vphantom{\sum}\right]\right), \label{eq7}
\end{equation}
where the complex amplitude $z(\omega,t)$ is such that $|z(\omega,t)|\leq1$.  Introducing the assumption \eqref{eq7} into \eqref{eq4}, one finds that $z(\omega,t)$ satisfies the Ott-Antonsen equation
\begin{equation}
\dot{z}(\omega,t)=i\omega z+(1-z^2)\frac{a}{2}+\frac{K}{2}Re^{-i\alpha}-\frac{K}{2}\overline{R}e^{i\alpha}z^2. \label{eq8}
\end{equation}
Quantity $z(\omega,t)$ should be interpreted as the frequency-dependent \emph{local order parameter}, in a sense that it quantifies the degree of synchrony of oscillators whose intrinsic frequencies $\omega_i$ lie within a small interval around the given frequency $\omega$. In the continuum limit, the global and the local order
parameter are connected by the self-consistency condition
\begin{equation}
R=\mathcal{G}z=\int_{-\infty}^{\infty}g(\omega)z(\omega)d\omega, \label{eq14}
\end{equation}
which follows from the definition \eqref{eq6} and the Ansatz \eqref{eq7}. Note that \eqref{eq8} presents a
generalization of the corresponding result in \cite{LCT10} for $a\neq 1,\alpha\neq 0$.

\section{Stationary solutions of the Ott-Antonsen equation}\label{sec3}

Within this section, our aim is to characterize the microscopic structure of the stationary solutions, finding the means to classify them by applying the self-consistency condition \eqref{eq14}. To do so, one first looks for the solutions of the Ott-Antonsen equation \eqref{eq8} for which the Kuramoto mean field $R(t)=\rho(t)e^{i\psi(t)}$ is constant. %Then we carry out the stability and bifurcation analysis of the macroscopic stationary states, explaining the scenarios that lead to the onset and the suppression of the collective mode.
In particular, we substitute the solution of the form $z(\omega,t)=r(\omega,t)e^{i\varphi(\omega,t)}$ into \eqref{eq8}, which ultimately results in
\begin{align}
\dot{r}&=\frac{B}{2}(1-r^2)\cos\phi, \nonumber \\
r\dot{\phi}&=\omega r-\frac{B}{2}(1+r^2)\sin\phi, \label{eq9}
\end{align}
having introduced the notation
\begin{align}
B&=\sqrt{a^2+K^2\rho^2+2aK\rho\cos(\psi-\alpha)}, \nonumber \\
\beta&=\arctan\frac{K\rho\sin(\psi-\alpha)}{a+K\rho\cos(\psi-\alpha)},\nonumber \\
\phi&=\varphi-\beta.\label{eq10}
\end{align}
From the system \eqref{eq9}, one infers that the quantity $B$, which depends only on the coupling strength and the mean field, plays the role of the \emph{macroscopic excitability parameter}. This follows from the fact that the microscopic structure of the stationary state is self-organized in a way that the assembly splits into two groups, according to the relation between the respective natural frequencies $\omega_i$ and $B$. In particular, one group is comprised of rotators in the \emph{excitable regime}, whose intrinsic frequencies satisfy $|\omega|<B$, whereas the other group consists of rotating units, whose intrinsic frequencies satisfy $|\omega|>B$. Another indication on the role of $B$ can be obtained if the definitions of $B$ and $\beta$ from \eqref{eq10} are applied to transform the original equation for the dynamics of rotators \eqref{eq1} into $\dot{\theta_i}=\omega_i-B\sin{(\theta_i-\beta)}$, which just conforms to a set of forced active rotators. From the level of single unit’s dynamics, $B$ is then classically referred to as the \emph{resistivity parameter}, in a sense that it reflects the rotator’s ability to modify its natural frequency.

Taking a closer look into the dynamics of the two subassemblies following from \eqref{eq9}, one finds that for $|\omega|<B$, there exist two steady states, given by
\begin{align}
r^*(\omega)&=1, \nonumber \\
\phi^*(\omega)&=\arcsin\frac{\omega}{B},\label{eq11}
\end{align}
and
\begin{align}
r^*(\omega)&=1, \nonumber \\
\phi^*(\omega)&=\pi-\arcsin\frac{\omega}{B},\label{eq12}
\end{align}
whereby our latter stability analysis will show that only the solution \eqref{eq11} is stable.
For the units within the rotating group $|\omega|>B$, the only steady state reads
\begin{align}
r^*(\omega)&=\frac{|\omega|}{B}-\sqrt{\frac{\omega^2}{B^2}-1} \nonumber \\
\phi^*(\omega)&=\frac{\pi}{2}\mbox{sgn}\omega. \label{eq13}
\end{align}
In order to fully quantify the stationary solutions of the Ott-Antonsen equation \eqref{eq8}, one has to obtain an explicit expression for the macroscopic excitability parameter $B$. In order to do so, we invoke the self-consistency equation \eqref{eq14}. Applying the latter to the stationary state $z^*(\omega)=r^*(\omega)e^{i\phi^*(\omega)+i\beta}$ given by \eqref{eq11} and \eqref{eq13}, one obtains
\begin{align}
\rho e^{i(\psi-\beta)}&=\frac{i\Omega}{B}+\int_{|\omega|<B}d\omega
g(\omega) \sqrt{1-\frac{\omega^2}{B^2}}-\nonumber\\
&-\frac{i}{B}\int_{|\omega|>B}d\omega
g(\omega)\omega\sqrt{1-\frac{B^2}{\omega^2}}, \label{eq15}
\end{align}
where $\Omega=\int_{-\infty}^{\infty}\omega g(\omega) d\omega$ refers to the mean value of the frequency distribution.
Separating for the real and the imaginary part of \eqref{eq15} and after some algebra, one ultimately arrives at the self-consistency equation for $B$ of the form:
\begin{align}
f(B)&=B^2-a^2-2K(f_1(B)\sin\alpha+f_2(B)\cos\alpha)+ \nonumber\\
&K^2\frac{f_1^2(B)+f_2^2(B)}{B^2}=0, \label{eq16}
\end{align}
where
\begin{align}
f_1(B)&=\Omega -\int_{|\omega|>B} d\omega g(\omega)\omega\sqrt{1-\frac{B^2}{\omega^2}}, \nonumber \\
f_2(B)&=\int_{|\omega|<B}d\omega g(\omega)\sqrt{B^2-\omega^2}. \label{eq17}
\end{align}
Note that the analogous expression has been obtained in \cite{LCT10}, but only for the particular case $a=1,\alpha=0$.
The results so far apply for an arbitrary distribution of natural frequencies $g(\omega)$. In order to carry out an explicit analysis on the stability of stationary states, including determining the associated stability boundaries and characterization of the transitions between the different collective regimes, we confine the remainder of the study to a particular case of $g(\omega)$, namely a \emph{uniform} distribution of frequencies on a bounded interval.

\section{Stability of the stationary solutions of the Ott-Antonsen equation}\label{sec4}

Within this Section, we specify the general results from Sec. \ref{sec3} to an example of a uniform distribution of natural frequencies $g(\omega)$ defined on an interval $\omega\in[\omega_1,\omega_2]$:
\begin{equation}
g(\omega)=\left\{
\begin{array}{ll}
0,&\;\omega<\omega_1,\\
\gamma,&\;\omega_1<\omega<\omega_2,\\
0,&\;\omega>\omega_2,\\
\end{array}
\right. \label{eq21}
\end{equation}
where $\gamma=1/(\omega_2-\omega_1)$ derives from the normalization condition. The given distribution is characterized by an average $\Omega=\frac{\omega_1+\omega_2}{2}$ and the width $\Delta=\omega_2-\omega_1$. The advantage of making such a choice of frequency distribution is that it allows for a full analytical treatment of the self-consistency equation \eqref{eq16} for the macroscopic excitability parameter. In particular, the integrals \eqref{eq17} then read
\begin{equation}
f_1(B)=\left\{
\begin{array}{ll}
\Omega-\gamma (F_1(\omega_2)-F_1(\omega_1)),&B<\omega_1,\\
\Omega-\gamma F_1(\omega_2),&\omega_1<B<\omega_2,\\
\Omega,&B>\omega_2,\\
\end{array}
\right. \label{eq22}
\end{equation}
where
\begin{equation}
F_1(\omega)=\frac{|\omega|}{2}\sqrt{\omega^2-B^2}+\frac{B^2}{2}\ln\frac{B}{|\omega|+\sqrt{\omega^2-B^2}},\label{eq23}
\end{equation}
and
\begin{equation}
f_2(B)=\left\{
\begin{array}{ll}
0,&B<\omega_1,\\
\gamma(\frac{\pi}{4} B^2-F_2(\omega_1)),&\omega_1<B<\omega_2,\\
\gamma(F_2(\omega_2)-F_2(\omega_1)),&B>\omega_2,\\
\end{array}
\right. \label{eq24}
\end{equation}
with
\begin{equation}
F_2(\omega)=\frac{|\omega|}{2}\sqrt{B^2-\omega^2}+\frac{B^2}{2}\arcsin\frac{\omega}{B}.\label{eq241}
\end{equation}

Considering the uniform frequency distribution \eqref{eq21}, we have carried out the stability and bifurcation analysis of the Ott-Antonsen equation \eqref{eq8}. The main control parameters are the characteristics of $g(\omega)$, namely its mean $\Omega$ and the width $\Delta$, while the remaining system parameters $a, K$ and $\alpha$ are kept fixed. Note that the stability analysis of \eqref{eq8} requires one to rewrite it as a real system in order to eliminate the complex conjugation \cite{WGO16,OW13,OW12}. The analysis \emph{per se} involves linearization of the Ott-Antonsen equation for variations around the stationary solution \eqref{eq11}-\eqref{eq13}, and consists in determining how the Lyapunov spectra of the stationary states depend on $\Omega$ and $\Delta$. While the technical details of the calculation are elaborated in the Appendix, the analysis we provide below will include characterization of the stationary solutions of the Ott-Antonsen equation \eqref{eq8} and the associated stability domains, as well as the description of the mechanisms behind the onset and the suppression of collective oscillations. The analytical results are corroborated by numerical experiments carried out on a heterogeneous assembly of $N=10^4$ active rotators.

\begin{figure}[t]%2
\centering
\includegraphics[scale=0.67]{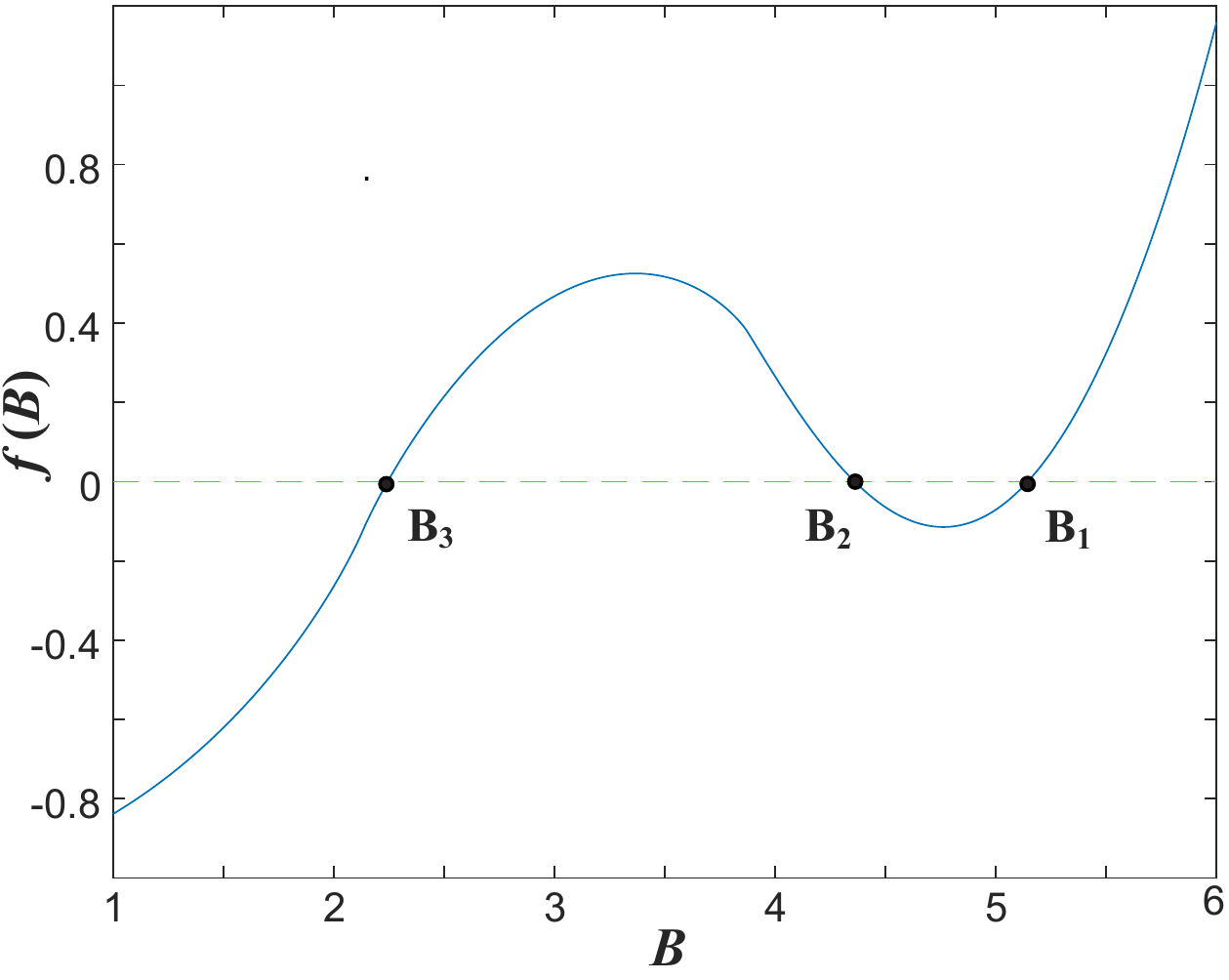}
\caption{Typical form of the function $f(B)$ and the three solutions $B_1>B_2>B_3$ of the self-consistency equation \eqref{eq16}. The system parameters are: $a=1$, $K =5$, $\alpha=0$, $\Omega=0.87$ and $\Delta=6$. }\label{fig1}
\end{figure}

The microscopic structure of the stationary regimes and the fashion in which their number and stability depend on the characteristics of $g(\omega)$ may conveniently be explained in terms of the solutions of the self-consistency equation \eqref{eq16} for the parameter $B$. A typical form of the function $f(B)$ for the considered domain of $(\Omega,\Delta)$ values is illustrated in Fig. \ref{fig1}. The three roots of $f(B)$, denoted by $B_1>B_2>B_3$, correspond to the stationary solutions of the Ott-Antonsen equation \eqref{eq8}. In particular, the macroscopic regime associated to $B_1$ presents a global rest state, because the macroscopic excitability parameter is so large that the frequencies of all the units lie below it. Given its microscopic structure, where the local dynamics is solely excitable, this state can also be termed a \emph{homogeneous stationary state}. The corresponding time series $\theta_i(t)$ and the evolution of the modulus of the Kuramoto order parameter $\rho(t)=|R(t)|$ are illustrated in Fig. \ref{fig3}(a). We shall demonstrate below that the global rest state may disappear in a fold bifurcation. In contrast to the macroscopic regime given by $B_1$, the stationary state corresponding to $B_3$ is typically a heterogeneous one, involving a subassembly of excitable units ($|\omega_i|<B_3$) and a subassembly of oscillating units ($|\omega_i|>B_3$), see the example of the time series in Fig. \ref{fig3}(c). In \cite{LCT10}, the heterogeneous stationary state is referred to as the asynchronous state, because spiking activity may be observed at the level of single units, but the macroscopic dynamics \emph{per se} does not exhibit a collective mode. The  heterogeneous state, as shown in greater detail below, may undergo either fold or Hopf bifurcation scenario. The stationary state associated to $B_2$ conforms to a saddle within the relevant $(\Omega,\Delta)$ domain, undergoing fold bifurcations either with $B_1$ or $B_3$, or providing for the separatrices in case of the two observed bistable regimes.

\begin{figure*}[t]%2
\centering
\includegraphics[scale=0.72]{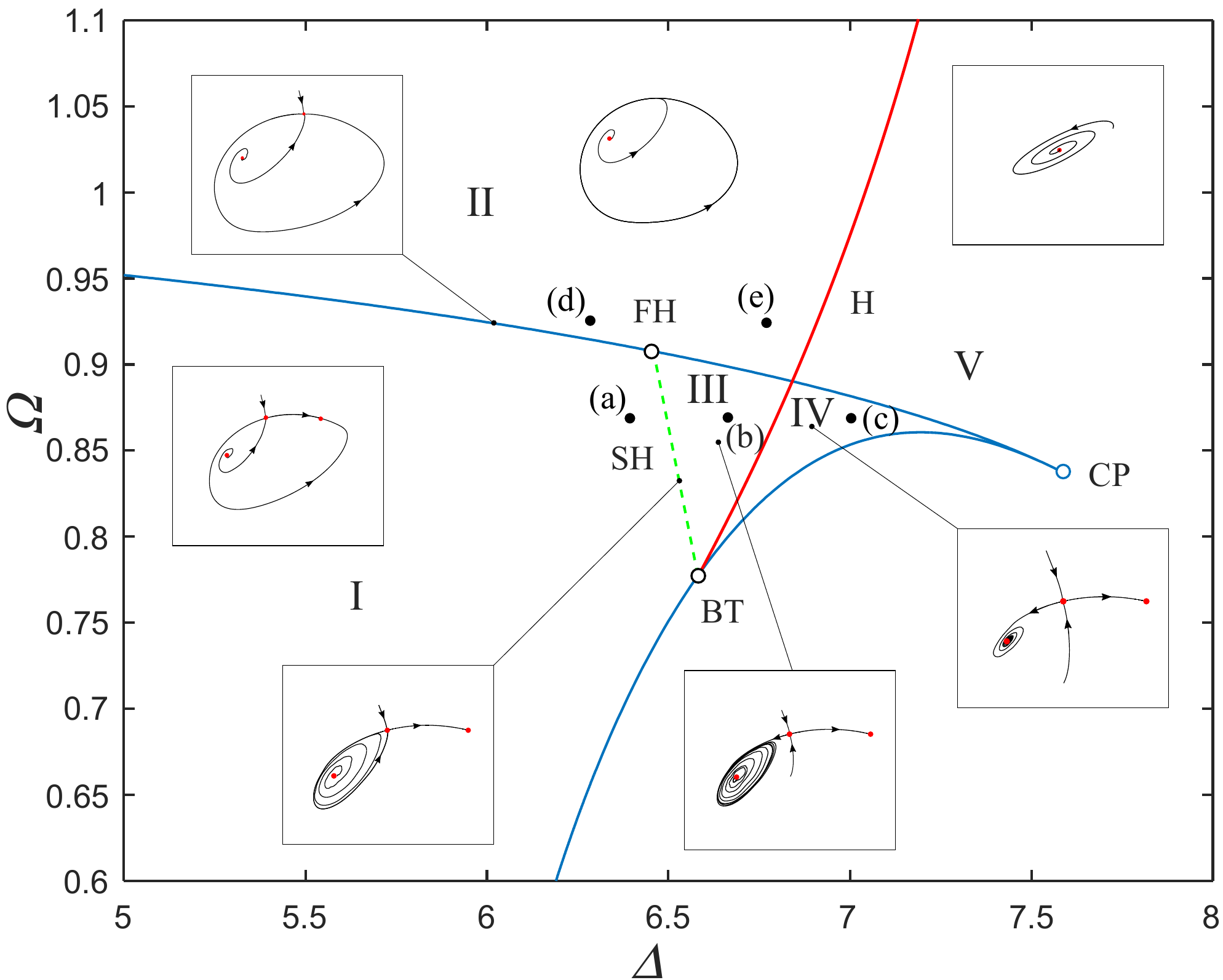}
\caption{Bifurcation diagram in the $(\Omega,\Delta)$ plane, constructed by the method of stability analysis described in the Appendix. The remaining system parameters are fixed to $a=1$, $K=5$, $\alpha=0$. The two branches of saddle-node bifurcations (blue solid lines) emanate from the cusp point CP, where the pitchfork bifurcation occurs. From the Bogdanov-Takens point (BT) emanate the Hopf bifurcation curve (H), indicated by the red solid line, and a branch of saddle-homoclinic bifurcations (SH), shown by the green dashed line. The upper branch of folds meets SH at the fold-homoclinic point (FH). The bullets indicate the parameter values associated to the time series in Fig. \ref{fig3}.}\label{fig2}
\end{figure*}

The bifurcation diagram in Fig. \ref{fig2} shows how the number and stability of the stationary solutions of the Ott-Antonsen equation \eqref{eq8} changes under variation of the parameters of the frequency distribution $\Omega$ and $\Delta$. The diagram features five characteristic domains I-V, and is organized around three codimension-2 bifurcation points, namely (i) the cusp point (CP), which corresponds to a symmetry-breaking pitchfork  bifurcation, (ii) the Bogdanov-Takens point (BT), which unfolds into Hopf (H) and saddle-homoclinic (SH) bifurcation curves, and (iii), the fold-homoclinic point (FH), where a branch of saddle-node bifurcations meets a curve of homoclinic tangencies of a limit cycle. The upper and the lower branch of folds, which emanate from the cusp, correspond to the coalescence of the state $B_2$ with $B_1$ and $B_3$, respectively. The former/latter branch has been obtained by solving for the parameters where the local minimum/maximum of the function $f(B)$ crosses the zero level. The Hopf bifurcation curve has been determined by the local stability analysis of the stationary state $B_3$. While such local analysis cannot provide for the saddle-homoclinic branch, its existence follows from the general structure of the Bodganov-Takens bifurcation \cite{SSK88,SZNS13}.

\begin{figure}[t]%2
\centering
\includegraphics[scale=0.49]{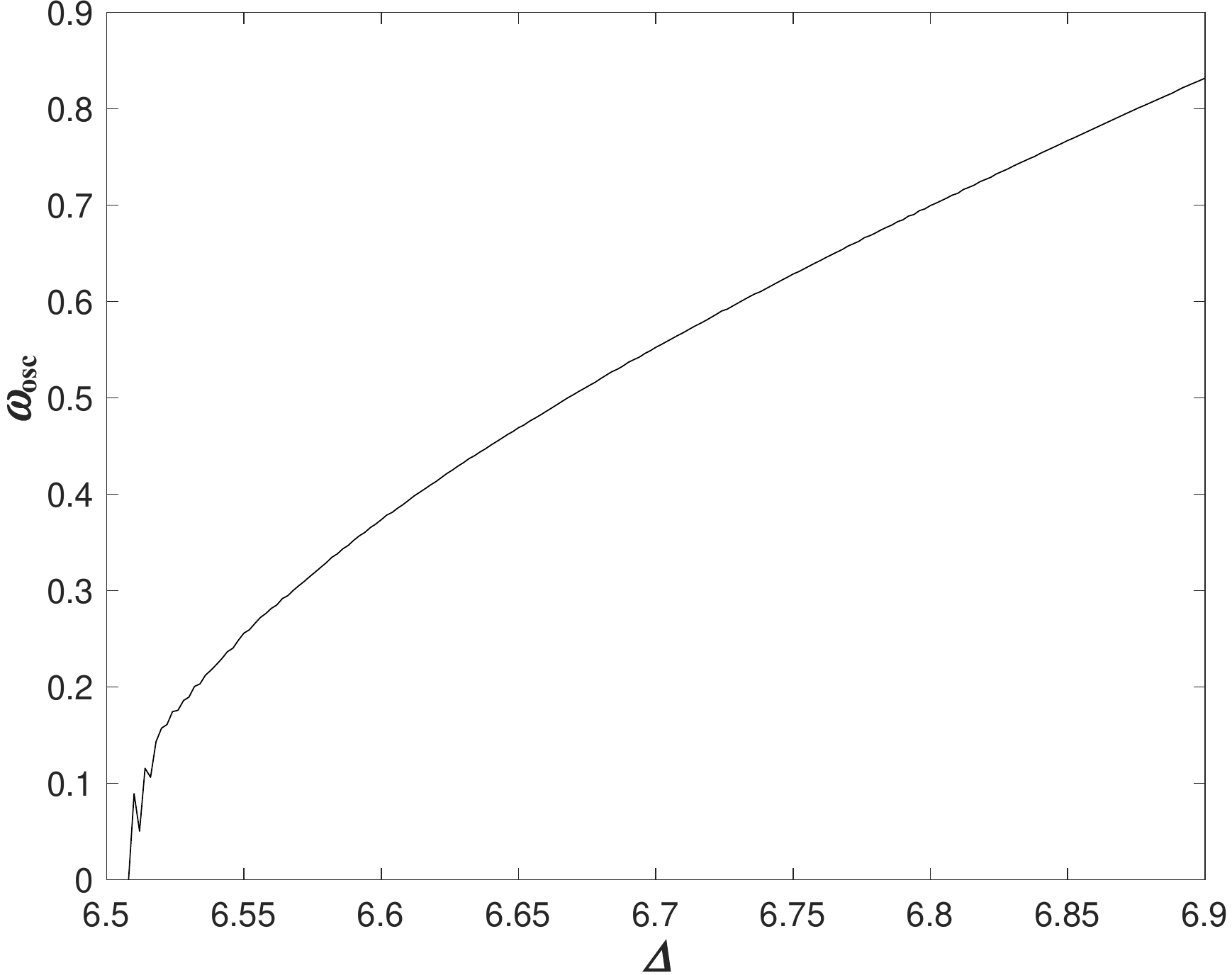}
\caption{Oscillation frequency of the periodic solution $\omega_{osc}$ in terms of diversity $\Delta$,
calculated along the Hopf bifurcation curve. One observes that the frequency tends to zero while approaching
the Bogdanov-Takens point. The parameters $a, K$ and $\alpha$ are the same as in Fig. \ref{fig2}.}\label{fig9}
\end{figure}

\begin{figure*}[t]%2
\centering
\includegraphics[scale=0.165]{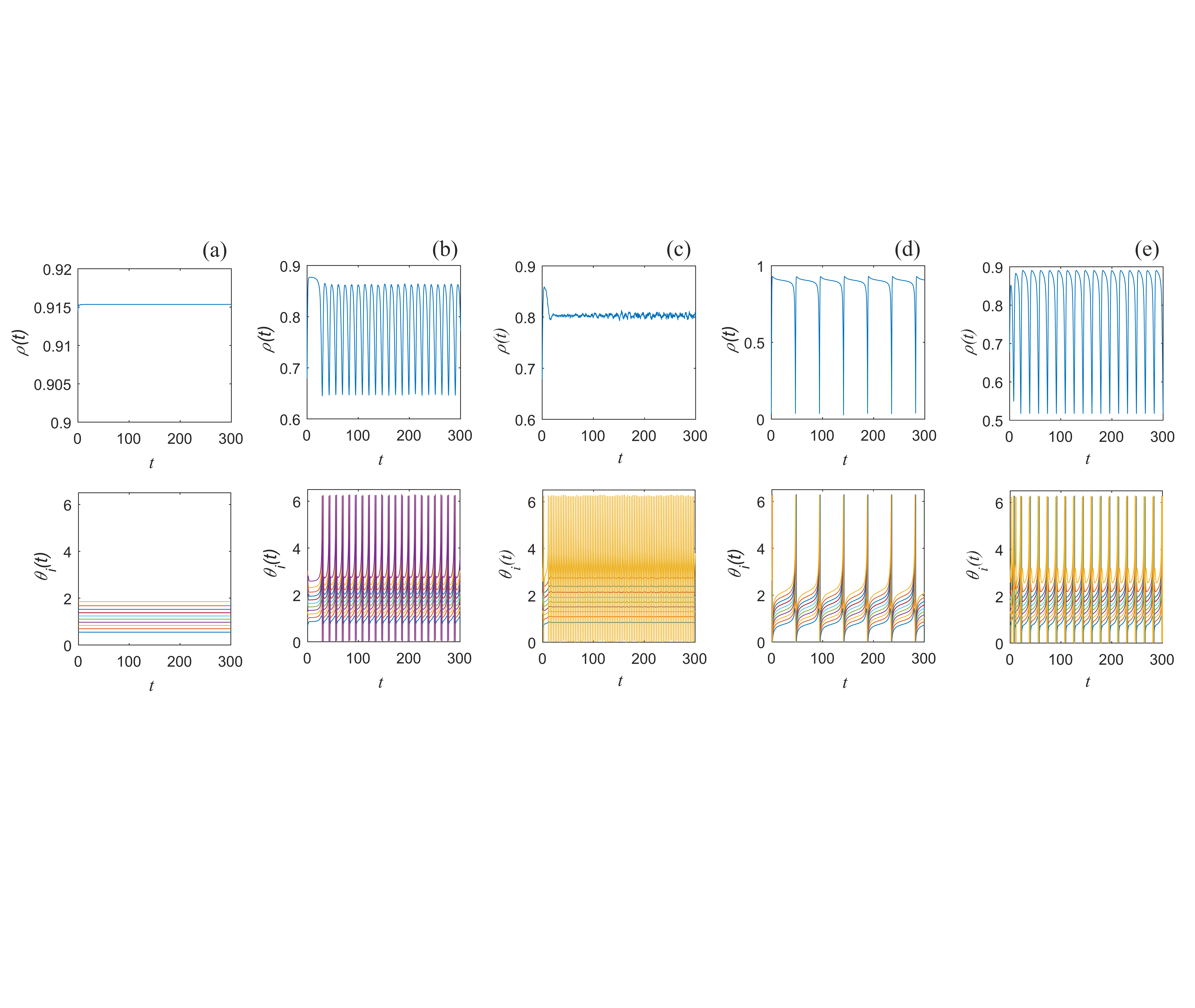}
\caption{Illustration of the local and the collective dynamics within the characteristic parameter domains indicated in Fig. \ref{fig2}. The top row shows the local time series $\theta_i(t)$ normalized over $2\pi$, while in the bottom row are provided the corresponding time series $\rho(t)=|R(t)|$. The particular parameter values of the frequency distribution (indicated by bullets in Fig. \ref{fig2}), are $(\Omega,\Delta)=(0.87,6.64)$ in (a), $(\Omega,\Delta)=(0.87,6.76)$ in (b), $(\Omega,\Delta)=(0.87,7)$ in (c), $(\Omega,\Delta)=(0.93,6.6)$ in (d) and $(\Omega,\Delta)=(0.93,6.78)$ in (e). The remaining system parameters are the same as in Fig. \ref{fig2}. }\label{fig3}
\end{figure*}

In the following, we provide a detailed description of the regimes underlying domains I-V, illustrating the associated phase portraits, cf. Fig. \ref{fig2}, and explaining the bifurcations that outline their stability boundaries.
At the cusp point CP, the two branches of saddle-node bifurcations coalesce, cf. the two blue solid lines in Fig. \ref{fig2}. In terms of the stationary states $B_1$-$B_3$ from Fig. \ref{fig1}, to the right of CP there exists only a stable fixed point B$_2$. Following the pitchfork bifurcation, $B_2$ becomes a saddle, whereas two stable nodes, $B_1$ and $B_3$, are created. The parameter region admitting only a single stable stationary state, be it $B_1, B_2$ or $B_3$ is denoted by V in Fig. \ref{fig2}. Decreasing the diversity, the stability of $B_1$ is influenced only by a fold bifurcation, whereas the character and stability of $B_3$ are influenced by the fold and Hopf bifurcations, derived from the Bogdanov-Takens point. We have evinced that while approaching BT, the frequency of oscillations $\omega_{osc}$ expectedly tends to zero, see Fig. \ref{fig9}. Along the lower branch of folds $B_2$ and $B_3$ get annihilated, so that from the right of this curve and to the cusp point, the only stable stationary state of the system is the node $B_1$. The Hopf bifurcation curve that emanates from the BT point affects the stability of the stationary state $B_3$, such that it becomes unstable for smaller diversities. This implies that within the region IV, bounded by the Hopf curve to the right and the two fold curves on the left, one observes \emph{bistability between two stationary states}, namely the stable node $B_1$ and the stable focus $B_3$, which are separated by the stable manifold of the saddle $B_2$, cf. the corresponding phase portrait in Fig. \ref{fig2}. Reducing diversity, $B_3$ undergoes a supercritical Hopf bifurcation (H), whereby immediately to the left of the Hopf curve (region III), one finds \emph{bistability between a small limit cycle and the stable node} $B_1$, again separated by the stable manifold of the saddle $B_2$. The time series illustrating the microscopic and macroscopic dynamics of the oscillatory states born from the Hopf bifurcation for two different parameter sets, $(\Omega_1,\Delta_1)=(0.87,6.76)$ and $(\Omega_2,\Delta_2)=(0.93,6.78)$, are provided in Fig. \ref{fig3}(b) and Fig. \ref{fig3}(e).

Consistent with the Bogdanov-Takens scenario, the limit cycle born from the Hopf bifurcation is destabilized via a homoclinic tangency to the saddle $B_2$, which is reflected by a branch of saddle-homoclinic bifurcations (SH) emanating from BT, see the green dashed line in Fig. \ref{fig2}. Using the local stability approach described in the Appendix, we are not able to trace the stability of a limit cycle \emph{per se}, but have been able to qualitatively verify the disappearance of the limit cycle by numerical means. The SH curve terminates at the fold-homoclinic point (FH), where it meets the upper branch of fold bifurcations. At FH, the stable manifold of the saddle $B_2$ touches the invariant circle. Decreasing diversity further away from the saddle-homoclinic bifurcation, cf. region I, the system exhibits a stable node $B_1$, and has two additional unstable fixed points, namely the saddle $B_2$ and the unstable focus $B_3$.

At the upper branch of folds, under increasing diversity, the stable node $B_1$ and the saddle $B_2$ collide and disappear. For $\Delta$ values less than that of the FH point, the fold takes place on the invariant circle, giving rise to a SNIPER bifurcation. Crossing the SNIPER bifurcation either by increasing $\Omega$ or $\Delta$, the collective dynamics of the system exhibits a transition toward the macroscopic oscillatory state. The latter is characterized by synchronous local oscillations of a large period, cf. the time series in Fig. \ref{fig3}(e). For this reason, it is also called the \emph{synchronous} state in \cite{LCT10}. For diversities to the right of the FH point, the saddle-node annihilation of $B_1$ and $B_2$ no longer occurs on an invariant circle. Thus, the only attractor within region VI corresponds to a small limit cycle emerging from Hopf destabilization of $B_3$.  For increasing diversity, $B_3$ gains stability by undergoing the inverse Hopf bifurcation, as already indicated above.

\subsection{Classical and hysteretic transitions between macroscopic regimes}\label{hyst}

\begin{figure*}[t]%2
\centering
\includegraphics[scale=0.19]{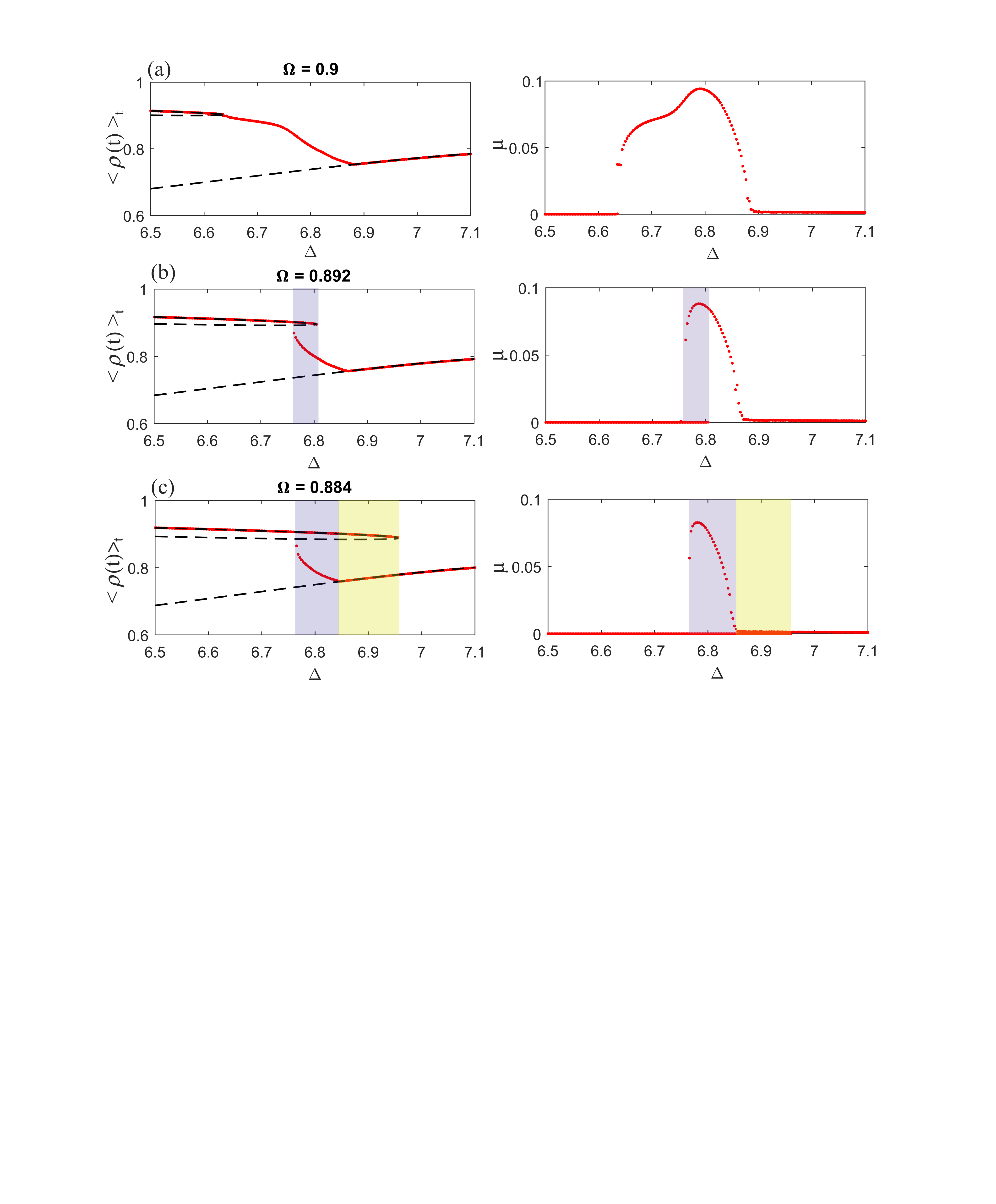}
\caption{Characteristic transition sequences between the different macroscopic regimes under increasing diversity for a fixed value of $\Omega$. The states are described by the time averaged modulus of the Kuramoto order parameter $\langle \rho(t)\rangle_t$ (left column) and the associated variance $\mu$ (right column). The mean frequencies are $\Omega=0.9$ in (a), $\Omega=0.892$ in (b) and $\Omega=0.884$ in (c). The classical scenario of transitions is recovered in (a), whereas the two hysteretic scenarios involving passage over one or two bistability regions, indicated by shading in (b) and (c), are reported for the first time.}\label{fig4}
%\vspace{-0.9cm}
\end{figure*}

Having characterized all the regimes of macroscopic activity and the associated stability domains, we focus on the scenarios leading to the onset and the suppression of the collective mode in heterogeneous populations, an issue of outstanding importance in the theory of coupled dynamical systems. By the classical paradigm \cite{LCT10}, the systematic increase of diversity under fixed mean frequency induces a sequence of transitions between the three
regimes of collective dynamics, namely the global rest state, the synchronous state (corresponding to macroscopic oscillations), and the asynchronous  state (a heterogeneous state displaying mixed excitable and oscillatory local dynamics). Our study demonstrates that apart from this, there exist two novel generic scenarios of transitions involving a \emph{hysteretic behavior}. To gain a deeper insight into this problem, we have plotted how the time-averaged modulus of the Kuramoto mean-field $\rho(t)=|R(t)|$ and the associated variance
$\mu=\sqrt{\langle \rho^2\rangle_t-\langle \rho\rangle_t ^2}$ change under variation of the diversity $\Delta$ for the three characteristic mean frequencies $\Omega\in\{0.9,0.892,0.884\}$, cf. Fig. \ref{fig4}. In order to reveal the potential bistable behavior, we have carried out sweeps in the directions of the increasing and the decreasing $\Delta$ applying the method of numerical continuation, where the initial conditions for the system with incremented $\Delta$ coincide with the final state at the previous $\Delta$ value.

The classical sequence of transitions is indeed recovered for $\Omega=0.9$, see Fig. \ref{fig4}(a). There, the onset of the collective mode is guided by a SNIPER bifurcation, mediating a transition from the homogeneous stationary state $B_1$ to a periodic solution. The suppression of the collective mode is induced by an inverse Hopf bifurcation that stabilizes the heterogeneous stationary state $B_3$, which is analogous to the Kuramoto-type scenario where the system desynchronizes under increasing disorder. For $\Omega=0.892$, we have established a hysteretic transition scenario, emerging due to a passage through a bistability region III from Fig. \ref{fig2}, which admits coexistence between the homogeneous stationary state $B_1$ and the periodic solution created from $B_3$, cf. Fig. \ref{fig4}(b). In this case, the onset of a collective mode is induced by a Hopf bifurcation, while its suppression is controlled by the homoclinic tangency of the limit cycle. For $\Omega=0.884$, the sequence of transitions remains hysteretic, but becomes more complex, see Fig. \ref{fig4}(c). In particular, by increasing the diversity, one traverses over two bistability regions, denoted by III and IV in Fig. \ref{fig2}. While the first one is qualitatively the same as for $\Omega=0.892$, the second one supports two coexisting stationary states, associated to $B_1$ and $B_3$. Nevertheless, the onset and the suppression of the collective mode \emph{per se} follow the same scenario as the one described in Fig. \ref{fig4}(b). Note that the described transition sequences are observed if the mean frequency $\Omega$ is
sufficiently large.

\begin{figure}[t]%2
\centering
\includegraphics[scale=0.495]{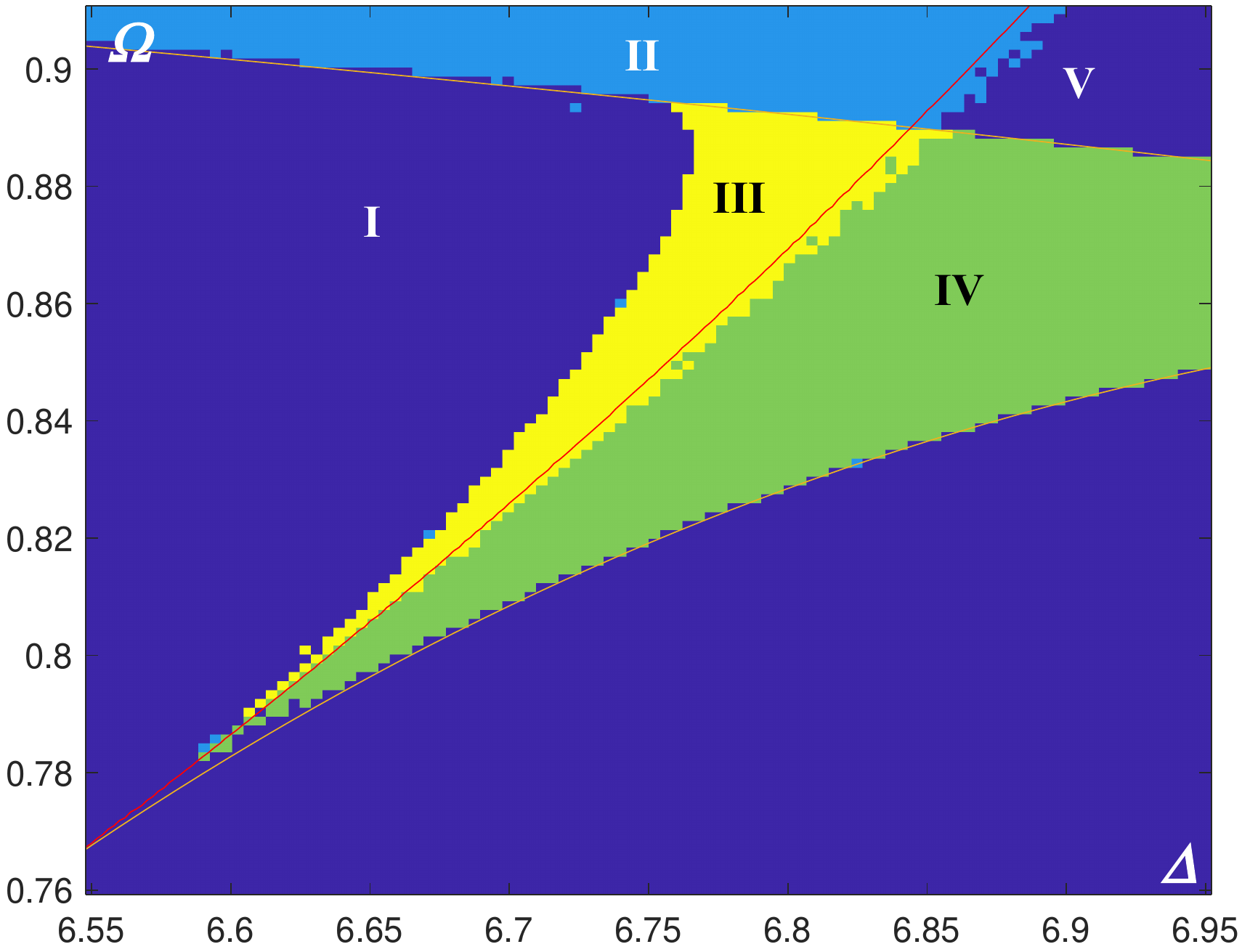}
\caption{The $(\Delta,\Omega)$ parameter plane divided into regions with different macroscopic dynamics: monostable stationary state (dark blue, regions I and V), monostable limit cycle (light blue, region II), bistability with two coexisting stationary states (green, region IV) and bistability between a stationary state and a limit cycle (yellow, region III). The parameter values are the same as in Fig. \ref{fig2}. Superimposed are the corresponding bifurcation curves obtained analytically within the Ott-Antonsen framework.}\label{fig5}
%\vspace{-0.9cm}
\end{figure}

In order to evince the generic character of the described scenarios and confirm the theoretical predictions regarding the parameter domains supporting the collective oscillations, we have carried out an extensive numerical study of the system's dynamics in terms of the parameters $\Delta$ and $\Omega$, see Fig. \ref{fig6}. In particular, using numerical continuation, we have performed bidirectional sweeps over the $(\Omega,\Delta)$ plane, keeping one of the parameters fixed while the other one was varied, in analogy to the method already described in relation to Fig. \ref{fig5}. This allowed us to partition the $(\Omega,\Delta)$ plane into different regions according to the number and the type of the supported attractors. Comparison of the boundaries of these regions with the bifurcation curves from Fig. \ref{fig2}, which are shown overlaid, corroborates an excellent agreement between the theory and the numerical results.

We have also examined whether the qualitative picture described so far persists under variation of the coupling strength $K$. It turns out that the general bifurcation structure holds qualitatively, which indicates the robustness of the scenarios underlying the transitions between the different collective regimes. Still, one notes that under increasing coupling strength, the cusp point and the Hopf bifurcation curve shift to a larger diversity (not shown).

\section{Impact of small coupling delay and small noise}\label{sec5}

In this section, the goal is to demonstrate that the physical picture described so far for the noiseless and the delay-free case qualitatively also holds in presence of \emph{small} noise or \emph{small} coupling delay. The small-noise scenario concerns a range of noise levels where the applied perturbation typically cannot give rise to noise-induced oscillations, but may rather evoke only rare spikes, so that the prevalent fraction of units within the excitable subassembly remains at the quasi-stationary state. The small-delay scenario refers to delay values which are significantly less than the typical period of local oscillations, such that no delay-induced oscillations or multistability can emerge \cite{YP09,KLSNY15,WYHS10}. Essentially, our intention is not to perform an exhaustive exploration of the effects of noise or coupling delay, but rather to confine the analysis to the cases where these two ingredients cannot evoke qualitatively new forms of collective behavior compared to the noiseless and delay-free case. We have carried out extensive numerical simulations to establish how the boundaries of the five characteristic domains in the $(\Omega,\Delta)$ plane are modified due to the action of small noise or small coupling delay.

\subsection{Effects of small coupling delay}

\begin{figure}[t]%2
\centering
\includegraphics[scale=0.15]{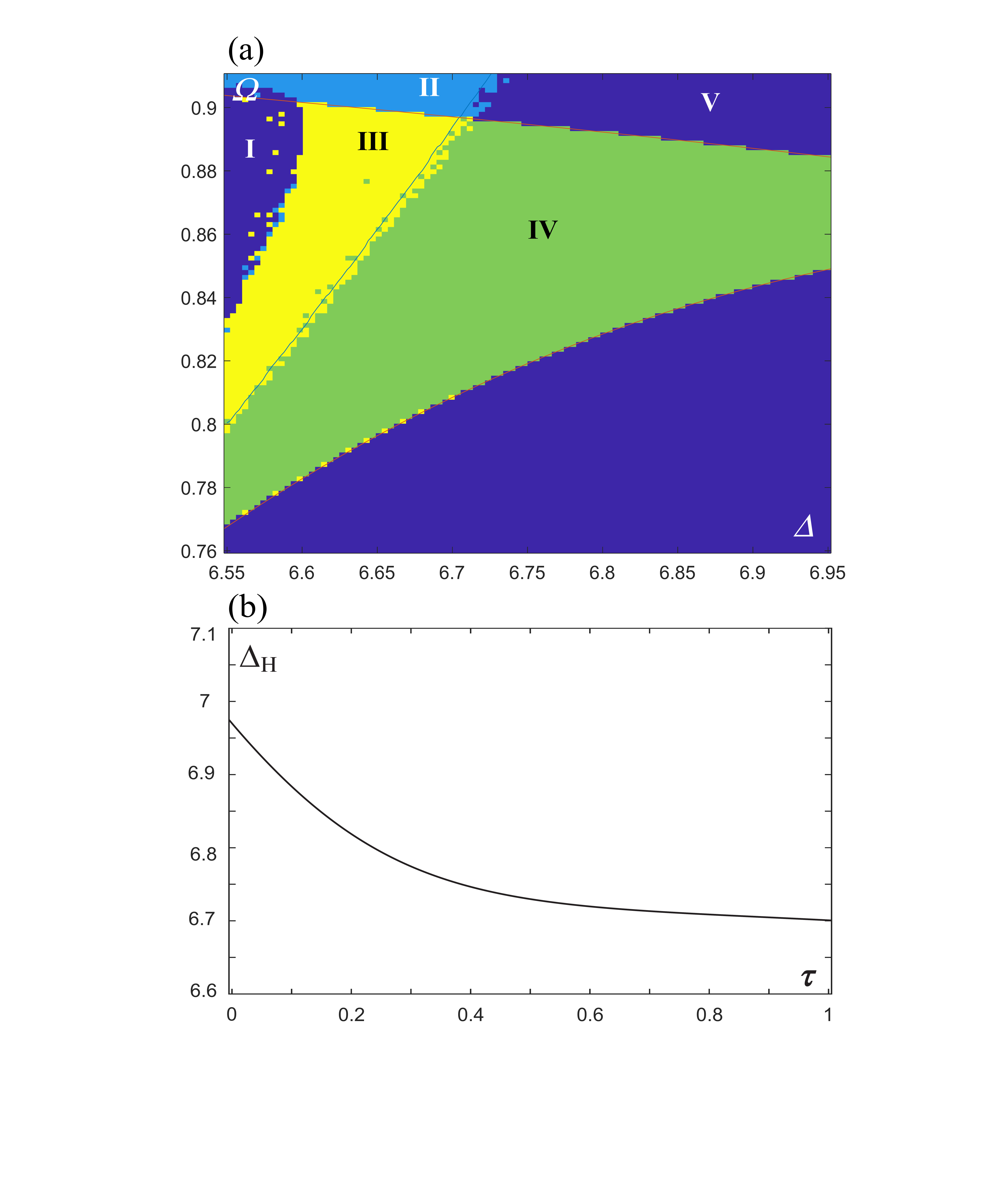}
\caption{(a) Characteristic domains of macroscopic behavior in the $(\Omega,\Delta)$ plane for coupling delay $\tau=0.3$. Color coding, as well as the remaining system parameters, are the same as in Fig. \ref{fig5}. Superimposed are the bifurcation curves obtained by the local stability approach described in the Appendix. (b) Critical diversity $\Delta_H$ corresponding to the Hopf destabilization of the state $B_3$ in dependence of $\tau$ for fixed $\Omega=0.88$.}\label{fig6}
%\vspace{-0.8cm}
\end{figure}

The effects of small coupling delay are illustrated in Fig. \ref{fig6}(a), which shows the characteristic domains of macroscopic behavior in the $(\Omega,\Delta)$ plane for the delay $\tau=0.3$. One observes an excellent agreement between the bifurcation curves, obtained analytically by the local stability approach described in the Appendix, and the associated stability boundaries of the domains. In particular, introducing the coupling delay does not affect the very coordinates of the stationary states of the Ott-Antonsen equation \eqref{eq8}, meaning that the branches of fold bifurcations remain unchanged relative to the delay-free case. Nevertheless, the key effect of the delay is that the Hopf bifurcation of the state $B_3$, which underlies one of the scenarios for the onset of the collective mode, shifts to a smaller diversity compared to the delay-free case. This implies that the delay promotes multistable behavior, in a sense that the bistability domain IV, characterized by the coexistence between the stable stationary states $B_1$ and $B_3$, becomes broader due to the impact of delay, cf. the green highlighted region in Fig. \ref{fig6}(a). From another point of view, the latter also suggests that the coupling delay promotes the onset of the collective mode via Hopf destabilization of the stationary state $B_3$, but suppresses the scenario where $B_1$ and $B_2$ undergo the SNIPER bifurcation. In Fig. \ref{fig6}(b) it is explicitly shown how the critical diversity $\Delta_H$ associated to Hopf bifurcation decreases with $\tau$ when $\Omega$ is kept fixed.

\subsection{Effects of small noise}

\begin{figure}[t]%2
\centering
\includegraphics[scale=0.635]{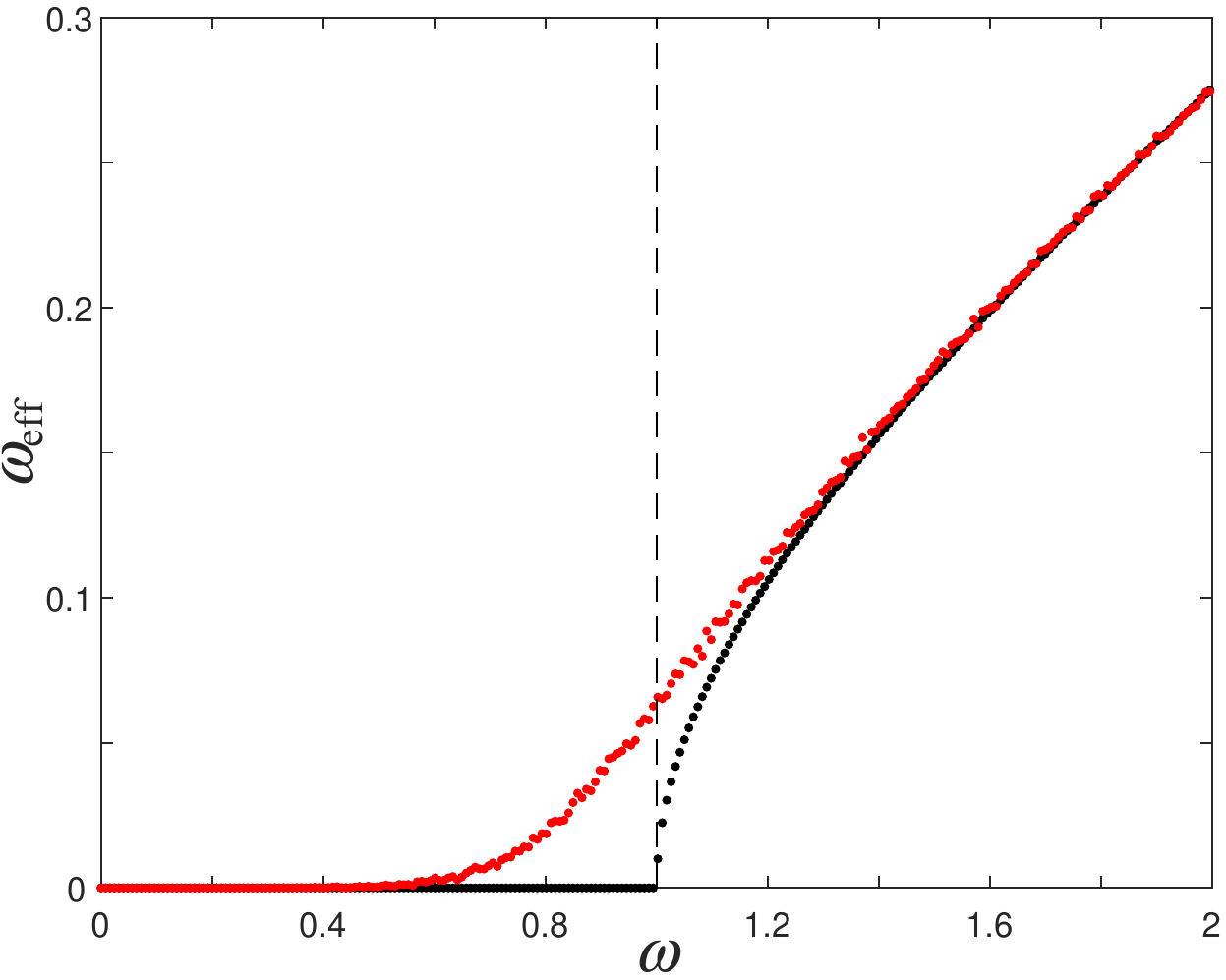}
\caption{Effective oscillation frequencies of uncoupled units $\omega_{eff}$ for the noiseless case (black dots) and under noise intensity $\sigma^2=0.09$ (red dots) as a function of the intrinsic parameters $\omega_i$. The dashed line indicates the excitability threshold $\omega=1$. The frequency distribution $g(\omega)$ is characterized by $\Omega=2, \Delta=4$.}\label{fig7}
%\vspace{-0.8cm}
\end{figure}
\begin{figure}[t]%2
\centering
\includegraphics[scale=0.137]{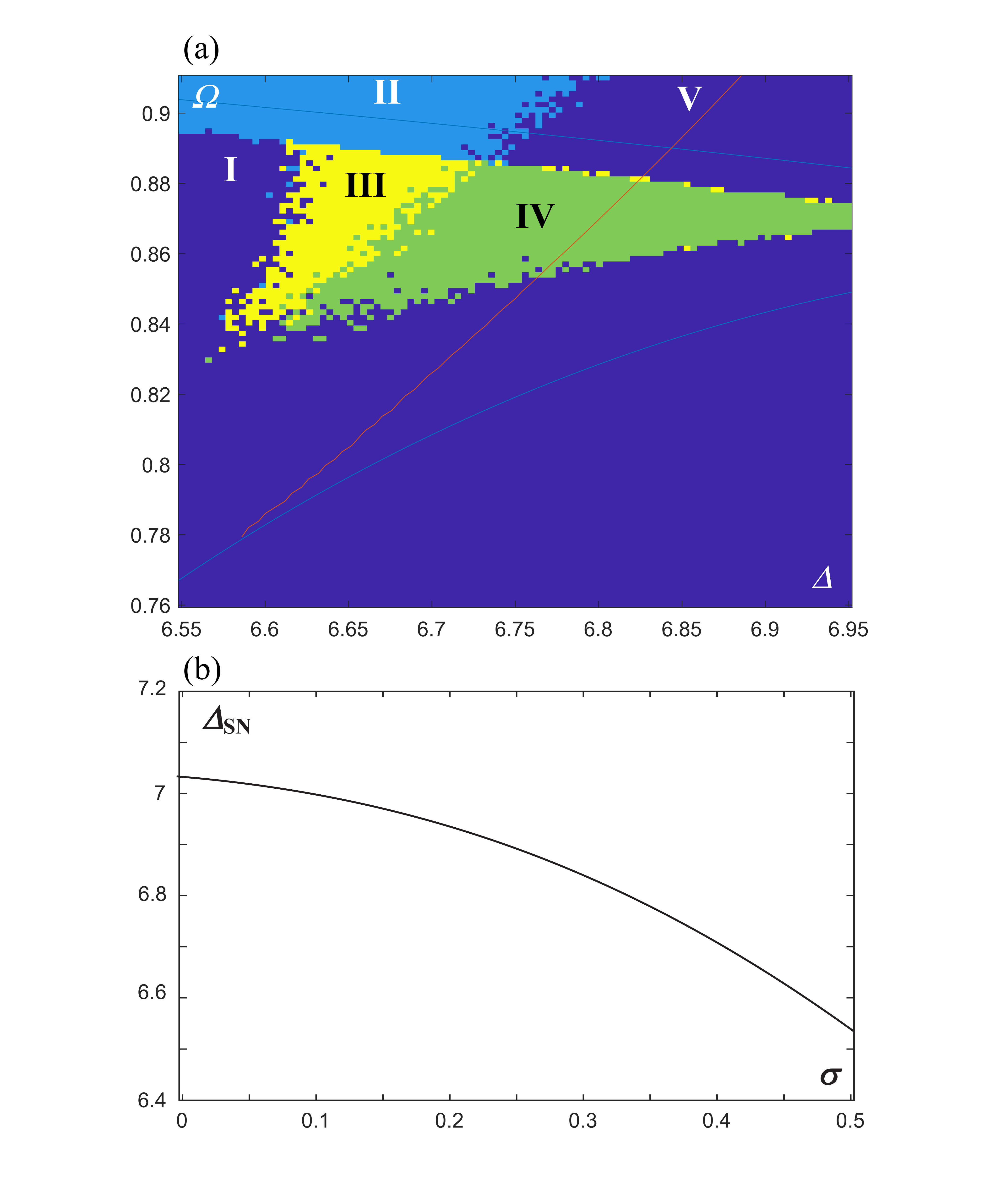}
\caption{(a) Characteristic domains of macroscopic dynamics in the $(\Omega,\Delta)$ plane for the noise level $\sigma=0.3$. The color coding and the remaining system parameters are the same as in Fig. \ref{fig5}. Superimposed are the bifurcation curves obtained analytically for the \emph{noise-free} case $\sigma=0$. (b) Decrease of the critical diversity $\Delta_{SN}$ with $\sigma$, corresponding to the saddle-node annihilation of the states $B_1$ and $B_2$ for fixed $\Omega=0.88$.}\label{fig8}
\end{figure}

In contrast to the impact of coupling delay, the small noise is found to influence the effective positions of both the fold and the Hopf bifurcation curves, cf. Fig. \ref{fig8}(a), where the five characteristic domains for the noise level $\sigma=0.3$ are shown together with the analytical curves for the \emph{noiseless} case. The primary effect of small noise is to promote the onset of the collective mode mediated via the SNIPER bifurcation, in a sense that for a fixed mean frequency $\Omega$, macroscopic oscillations can be observed for the diversity $\Delta$ smaller than those in the noiseless case. As a consequence, one observes that the critical diversity $\Delta_{SN}$ at which the fold between the states $B_1$ and $B_2$ takes place reduces under increasing $\sigma$, as indeed shown in Fig. \ref{fig8}(b) for the fixed $\Omega=0.88$. Nonetheless, noise also shifts the location of the Hopf bifurcation relevant for the stability of the state $B_3$, see Fig. \ref{fig8}(a). This may be interpreted as a disordering effect of noise, in a sense that the transition from the regime of macroscopic oscillations (domain II) to the asynchronous regime (domain V) occurs at the diversity smaller than that for the noise-free case. Also note that the bistability regions III and IV shrink as compared to the noiseless case.

In principle, one observes that the structure of the characteristic domains is qualitatively preserved with introduction of small noise, but the associated stability boundaries shift to the left with respect to the noiseless case. This can be understood by the following qualitative reasoning. The impact of small noise on the local dynamics of the nodes can roughly be interpreted as a perturbation of the intrinsic frequency $\omega_i$. To corroborate this, in Fig. \ref{fig7} we illustrate how the \emph{effective} oscillation frequencies of single units $\omega_{eff,i}$, calculated numerically as the inverse of the respective mean oscillation periods, change in presence of noise $\sigma=0.3$. One finds that a certain fraction of units whose intrinsic frequencies $\omega_i$ lie closest to the excitability threshold $\omega=1$ acquire a non-zero effective frequency, i.e. manifest noise-induced oscillations, while the excitable units further away from the threshold remain quasi-stationary. Nonetheless, the impact of noise on the self-oscillating units is reflected as a small increase of their effective frequency. Thus, in qualitative terms, the effect of small noise amounts to enhancing the effective frequency of the units near the threshold $\omega=1$. Since this effect is symmetrical for positive and negative $\omega$, the average assembly frequency $\Omega$ remains unchanged, whereas the variance of the associated distribution increases proportionally to the noise intensity. Therefore the introduction of small noise should lead to similar effects as the increase of diversity $\Delta$.

\section{Summary and Conclusion}\label{sec6}

Considering a heterogeneous assembly of active rotators displaying excitable or oscillatory local dynamics, we have classified the associated macroscopic regimes and have demonstrated the generic scenarios for the onset and the suppression of collective oscillations. The analytical part of the study has been carried out within the framework of Ott-Antonsen theory applied for the delay- and noise-free system in the continuum limit, which enabled us to determine the three macroscopic stationary states in case of an arbitrary distribution of natural frequencies. The main qualitative insight into the microscopic structure of stationary states is that the population may in principle split into the excitable and the rotating subassembly, with the division depending on the relationship between the respective natural frequency of a rotator and the macroscopic excitability parameter. In this context, we have identified a homogeneous equilibrium where the units typically lie at rest, as well as a heterogeneous (mixed) collective stationary state, comprised of units either in the excitable or the oscillatory regime. The local approach to stability and bifurcation analysis of the stationary states we have derived allowed us to address both the delay-free case and the case where the system’s behavior is influenced by coupling delay. The analysis has been specified to the particular case of a uniform frequency distribution on a bounded interval. While the stationary states have been determined earlier for a similar, but a less general model \cite{LCT10}, the stability analysis, as presented here, has been carried out for the first time.

We have demonstrated that the complex bifurcation structure underlying the stability boundaries of the different macroscopic regimes is organized by three codimension-two bifurcation points, including the Bogdanov-Takens point, the cusp point and the fold-homoclinic point. Our analysis has revealed the existence of five characteristic domains, three of which support the monostable collective behavior, while two admit bistability, involving either the coexistence between two stable stationary states, or the coexistence between a stationary and a periodic solution. We have found that depending on the mean frequency, the onset and the suppression of the collective mode may emerge via two qualitatively different scenarios under variation of diversity. In particular, for a smaller mean frequency, the onset of collective oscillations under decreasing diversity occurs due to a Hopf destabilization of a stationary state, whereas the oscillations are terminated via a saddle-homoclinic bifurcation. Nevertheless, for a sufficiently large mean frequency, increasing the diversity gives rise to collective oscillations in a SNIPER bifurcation, while the suppression of oscillations is due to an inverse Hopf bifurcation.

The classical paradigm concerning the sequence of transitions between the collective regimes in heterogeneous systems under increasing diversity involves three characteristic states, namely the global rest state, the synchronous state, characterized by macroscopic oscillations, and the asynchronous state, based on mixed excitable and oscillatory local dynamics \cite{LCT10}. In addition to this paradigm, our analysis has revealed two novel scenarios, which are hysteretic, and involve a passage through one or two bistable domains. By the first scenario, the transition from the global rest state to the asynchronous state occurs via two bistable regimes, the first involving a   coexistence between a periodic solution and the rest state, and the second one, featuring coexistence between the rest state and the asynchronous state. The second hysteretic scenario is similar, but the intermediate stage involves only the coexistence between the homogeneous and the oscillatory state.

Combining theoretical methods and numerical experiments, we have shown that the basic bifurcation structure from the delay- and noiseless case persists in presence of small noise or small coupling delay. Nevertheless, these two ingredients are found to modify the stability boundaries of the five characteristic domains. In particular, due to coupling delay, the position of the Hopf bifurcation curve is shifted toward the smaller diversity, which effectively promotes the Hopf-mediated onset of macroscopic oscillations, and also enhances the parameter domain supporting bistability. Noise is seen to affect both the fold and the Hopf bifurcations, whereby the effective position of the fold/Hopf curve is shifted to smaller mean frequency/smaller diversity. At the level of macroscopic behavior, this is reflected as the promotion/suppression of the onset of macroscopic oscillations via SNIPER/Hopf bifurcation scenario, contributing in addition to a reduction of the two bistability domains. While the described bifurcation structure appears to be generic for the considered type of frequency distribution, remaining qualitatively similar under the influence of small noise or small coupling delay, it would be interesting to examine whether and how it is modified for a substantially different form of a frequency distribution, such as a bimodal one.

\appendix*
\section{Calculation of the stability of the stationary solution of the Ott-Antonsen equation} \label{app1}

Here we elaborate on the method applied to calculate the stability of the stationary solutions of the Ott-Antonsen equation \eqref{eq8}. In particular, we first introduce the expressions $z(\omega,t)=x(\omega,t)+iy(\omega,t)$ and $R(\omega,t)=X(\omega,t)+iY(\omega,t)$ for the local and the global order parameter, respectively, transforming \eqref{eq8} to
\begin{align}
\dot{x}&=F(x,y,X,Y)=\frac{a}{2}(y^2-x^2+1)-\omega y-\nonumber\\
&-Kxy(Y\cos\alpha -X\sin\alpha)-
\frac{K}{2}(X\cos\alpha+Y\sin\alpha)\cdot \nonumber\\
&\cdot(x^2-y^2)+\frac{K}{2}(X\cos\alpha+Y\sin\alpha)\nonumber\\
\dot{y}&=G(x,y,X,Y)=-axy+\omega x-Kxy(Y\sin\alpha+X\cos\alpha)+\nonumber\\
&+\frac{K}{2}(Y\cos\alpha- X\sin\alpha)(x^2 - y^2)+\nonumber\\
&+\frac{K}{2}(Y\cos\alpha-X\sin\alpha). \label{eq25}
\end{align}

The linearization of Ott-Antonsen equation \eqref{eq8} for variations $\xi=(\delta x,\delta y)^T,\Xi=(\delta X,\delta Y)^T$ of the stationary solution $(x_0,y_0)$ can then succinctly be written in the matrix form as
\begin{equation}
\frac{d\xi(\omega,t)}{dt}=A(\omega)\xi(\omega,t)+B(\omega)\Xi(t), \label{eq26}
\end{equation}
where the matrices of derivatives are
\begin{equation}
A(\omega)=\left(\begin{array}{ll}
\frac{\partial F}{\partial x}&\frac{\partial F}{\partial y}\\
\frac{\partial G}{\partial x}&\frac{\partial G}{\partial y}\\
\end{array}\right),\;B(\omega)=\left(\begin{array}{ll}
\frac{\partial F}{\partial X}&\frac{\partial F}{\partial Y}\\
\frac{\partial G}{\partial X}&\frac{\partial G}{\partial Y}\\
\end{array}\right). \label{eq27}
\end{equation}
Assuming that the variation $\xi(\omega,t)$ satisfies the Ansatz $\xi(\omega,t)=\xi(\omega)e^{\lambda t}$, and similarly $\Xi(t)=\Xi e^{\lambda t}$, \eqref{eq26} becomes
\begin{equation}
(A(\omega)-\lambda I)\xi(\omega)+B(\omega)\Xi=0, \label{eq27}
\end{equation}
where $I$ denotes the identity matrix. As shown in \cite{OW13}, the continuous Lyapunov spectrum consists of the eigenvalues of the matrix $B(\omega)$ for all $\omega\in[\omega_1,\omega_2]$. In our case, the continuous spectrum turns out to be always stable or marginally stable, such that the stability of the stationary solutions is determined by the discrete spectrum. In order to obtain the discrete spectrum, we multiply \eqref{eq27} from the left by $g(\omega)(A(\omega)-\lambda I)^{-1}$ and integrate over $\omega$ obtaining $C(\lambda)\Xi=0$, where
\begin{equation}
C(\lambda)=I+\int_{-\infty}^{\infty}d\omega g(\omega)(A(\omega)-\lambda I)^{-1}B(\omega). \label{eq28}
\end{equation}
The discrete Lyapunov spectrum can then be calculated by numerically solving the system $\det C(\lambda)=0$.

In the case of non-zero coupling delay, the same type of analysis remains valid, while one has to replace $X$ and $Y$ in the r.h.s. of \eqref{eq25} by their delayed counterparts $X(t-\tau)$ and $Y(t-\tau)$. This leads to the same matrix $C(\lambda)$ as in \eqref{eq28}, with the only difference being the substitution of $B(\omega)$ by $B(\omega)e^{-\lambda\tau}$.

\begin{acknowledgments}
The work on Sections III and IV was supported by the Russian Foundation for Basic Research under project No. 17-02-00904. The work on Section V was supported by the Russian Foundation for Basic Research under project No. 19-52-10004. The numerical simulations were supported by the Russian Science Foundation under project No. 19-72-10114. IF acknowledges the support from the Ministry of Education, Science and Technological Development of the Republic of Serbia under project No. 171017. The authors would like to thank Matthias Wolfrum for fruitful discussions during the various stages of the study.
\end{acknowledgments}

\end{document}